\documentclass[aps,pre,nofootinbib,amsmath,amsfonts]{revtex4}
\usepackage{hyperref}
\usepackage{graphicx,xcolor}
\usepackage{longtable}
\usepackage{mathtools}

\newcommand{\eqr}[1]{Eq.~\eqref{#1}}

\newcommand{\secr}[1]{Sec.~[\ref{#1}]}

\newcommand{\tblr}[1]{Table~\ref{#1}}
\newcommand{\figr}[1]{Fig.~\ref{#1}}

\newcommand{\velocity}{{\rm{v}}}

\newcommand{\vR}{{\mathbf{r}}}

\newcommand{\bbeta}{{\scriptscriptstyle\beta}}
\newcommand{\scaleprofile}{0.9}
\newcommand{\scaleradius}{0.9}
\newcommand{\excess}[1]{\mathfrak{E}\left[\,#1\,\right]}

\begin{document}

\title{Curvature dependence of the interfacial heat and mass transfer coefficients.}
\author{K.~S.~Glavatskiy$^{1,2}$}
\author{D.~Bedeaux$^{2}$}
\affiliation{$^{1}$ School of Applied Sciences, Royal Melbourne Institute of Technology, Melbourne, VIC 3001, Australia}
\affiliation{$^{2}$ Department of Chemistry, NO 7491, Norwegian University of Science and Technology, Trondheim, Norway}

\begin{abstract}
Nucleation is often accompanied by heat transfer between the surroundings and a nucleus of a new phase. The interface between two phases gives an additional resistance to this transfer. For small nuclei the interfacial curvature is high, which affects not only equilibrium quantities such as surface tension, but also the transport properties. In particular, high curvature affects the interfacial resistance to heat and mass transfer. We develop a framework for determining the curvature dependence of the interfacial heat and mass transfer resistances. We determine the interfacial resistances as a function of a curvature. The analysis is performed for a bubble of a one-component fluid and may be extended to various nuclei of multicomponent systems. The curvature dependence of the interfacial resistances is important in modeling transport processes in multiphase systems.
\end{abstract}

\maketitle



\section{Introduction.}

Mesoscale structures in soft matter can spontaneously form in such systems as surfactant solutions \cite{Komura2007}. They are characterized by small aggregates of a new phase on a nanometer scale. The growth of nuclei is the first step in a macroscopic phase transformation \cite{phasetransitions}. These processes have been studied more than a hundred years \cite{Feder1966}. During nucleation, there is an energy barrier due to the energy costs to create an interface between the two phases \cite{Kashchiev}. The classical theory of nucleation fails to describe the results of experiments adequately \cite{Vehkamaki}. A number of extensions have therefore been proposed \cite{Reguera2004}. Some employ kinetic equations \cite{Dubrovskii2010}, other molecular dynamic simulations \cite{Kathmann2009}. The overall picture is still far from clear \cite{Lervik2009}. The overall reason for a system to nucleate, however, is to decrease the total system's Gibbs energy. The homogeneous phase inside the nucleus has a lower Gibbs energy than the nucleating phase. Below this critical size, the nucleus prefers to shrink. Above this critical size, it starts to grow. The typical size of the critical nucleus is of the order of nanometers. An important aspects of the nucleation process is a small radius of a nucleus, which corresponds to its high curvature. The curvature increases the surface tension relative to the surface tension of a flat interface \cite{Helfrich1973}. 

One of the main issues, which make nucleation complicated, is that it is a non-equilibrium process \cite{Reguera2004}. There exist fluxes of heat and mass between the surroundings an the nuclei, which facilitate its growth. The interface between two phases has an additional resistance for heat and mass transfer even for a flat interface \cite{Glavatskiy2010b}. During nucleation the interfacial curvature is finite and changes the heat and mass resistances of the interface. It is the aim of this paper to investigate how the curvature of a small nucleus affects the interfacial resistance.

Earlier we developed a tool to analyze the interfacial resistances for flat interfaces \cite{Glavatskiy2010a, bedeaux/vdW/IV}. We introduced integral relations which allowed us to calculate the interfacial resistances knowing only the equilibrium properties of the system. This is very useful for our purpose, since nucleation is a non-stationary process, which is complicated to analyze. Earlier we calculated the resistances of flat surfaces \cite{Glavatskiy2010b} and in this paper we will extend the analysis to a spherical surface.

The equilibrium properties of the system are modeled with the help of the square gradient model \cite{Yang/surface, vdW/translation}. The square gradient theory for the interfacial region originates from the work of van der Waals \cite{vdW/sg} for liquid-vapor equilibrium of a one-component fluid and the work of Cahn and Hilliard \cite{cahnhilliard/fens/I} for fluid-fluid equilibrium of binary mixtures. The introduction of the gradients of the densities in the thermodynamic description successfully explains macroscopic thermodynamic behavior of two-phase coexistence, in particular, the surface tension \cite{Yang/surface}. It has been widely used to model the surface behavior of planar fluid interfaces \cite{Lamorgese2011}. A systematic extension of this theory to non-equilibrium systems using non-equilibrium thermodynamics in two-phase multi-component systems has been given \cite{GlavSpringer, bedeaux/vdW/I, bedeaux/vdW/II, bedeaux/vdW/III, Glavatskiy2008, Glavatskiy2009}. 

In this paper we establish a method to calculate the interfacial resistances for spherical interfaces. The results can be in principle verified in molecular simulations \cite{Lervik2009} or experiments. Performing a particular measurement of the interfacial resistance may be a complicated process, so it is important to understand what data one may expect from particular experiments. Here we consider a spherical bubble of a one-component system. However, the analysis is applicable to droplets and multicomponent systems with no restrictions in generality. The paper is organized as following. In \secr{sec/Gradient} we give a brief description of the key points of the square gradient model. In \secr{sec/Excess} we discuss how the macroscopic properties of the system are connected to the local continuous profiles. We introduce the excess densities and excess resistances, the properties which describe behavior of the entire interface. In \secr{sec/Results} we present calculations of the interfacial resistances according to the developed model. The curvature dependence of the resistances is discussed. Finally, in \secr{sec/Conclusion} we summarize our findings.

\section{Local description of a spherical surface.}\label{sec/Gradient}

For a one-component system the specific local free energy in the interfacial region is a function of both the mass density $\rho(\vR)$ and the mass density gradient $\nabla{\rho(\vR)}$:
\begin{equation}\label{eq/Gradient/Eq/01}
f(\vR) = f_{0}(T,\,\rho) + {\frac{1}{2}}\frac{\kappa}{\rho(\vR)}\,|\nabla{\rho(\vR)}|^{2} 
\end{equation}%
where $f_{0}$ is the homogeneous free energy and $\kappa$ is a parameter of the model, independent of the temperature, which should be chosen such that the value of the surface tension reproduces a typical experimental value. In equilibrium the total Helmholtz energy reaches its minimum given the condition that the total mass is fixed. This requires  minimization with respect to the density of the grand potential $\Omega = - \int{d\vR\,p(\vR)}$, where $p\,(\vR) = (\mu_{e} - f(\vR))\rho(\vR)$ and $\mu_{e}$ is the equilibrium chemical potential:
\begin{equation}\label{eq/Gradient/Eq/03}
\mu_{e} = {\frac{\partial(\rho f_{0})}{\partial \rho}}  - \kappa\Delta\rho \\%
\end{equation}%
where $\Delta \equiv \nabla\cdot\nabla$ is the Laplace operator. In a spherically symmetric system all the quantities depend only on the radial coordinate so that \eqr{eq/Gradient/Eq/03} becomes
\begin{equation}\label{eq/Gradient/Eq/07}
\mu_{e} = \mu_{0}(\rho, T) - \kappa\left(\rho^{\prime\prime} + \frac{2}{r}\rho^{\prime}\right) \\%
\end{equation}%
where $\mu_{0}(\rho, T)$ is the homogeneous chemical potential and prime indicates derivative with respect to the radius. The actual density profile can be found from \eqr{eq/Gradient/Eq/07}.

In the interfacial region the pressure becomes a tensor 
\begin{equation}\label{eq/Gradient/Eq/05}
\sigma_{\alpha\beta}(\vR) = p\,(\vR)\,\delta_{\alpha\bbeta} + \gamma_{\alpha\bbeta}(\vR)
\end{equation}%
where $\gamma_{\alpha\bbeta}(\vR) \equiv \kappa\,\nabla_{\alpha}\rho(\vR)\nabla_{\bbeta}\rho(\vR)$. In a spherically symmetric system $\sigma_{\alpha\beta}$ has a diagonal form with $\sigma_{11}(r) \equiv p_{n}(r)$ being the so-called normal pressure
\begin{equation}\label{eq/Gradient/Eq/08}
p_{n}(r) = p\,(r) + \gamma_{rr}(r) = p\,(r) + \kappa {\rho^{\prime}(r)}^{2}
\end{equation}%
and $\sigma_{22}(r) = \sigma_{33}(r) \equiv p_{t}(r)$ being the tangential pressure: $p_{t}(r) = p\,(r)$. 

Note that, unlike in the system with planer interface, the normal pressure is not constant with respect to the position through the interface. This leads to the existence of Laplace pressure \cite{Yang/surface}.

In non-equilibrium the thermodynamic properties change not only with position, but also with time. Furthermore, the temperature and the chemical potential are no longer constant. However, a local description of the interfacial region can still be given with the help of the square gradient model. To extend the equilibrium square gradient model to non-equilibrium we will assume that all the thermodynamic densities are given by the same expressions as in equilibrium \cite{Glavatskiy2009}. In particular the specific Helmholtz energy 
\begin{equation} \label{eq/Non-equilibrium/01}
f(r, t) = \mu(r, t) - p\,(r, t)/\rho(r, t)
\end{equation}
Furthermore, the chemical potential 
\begin{equation}\label{eq/Non-equilibrium/02}
\mu(r, t) = \mu_{0}(\rho(r, t), T(r, t)) - \kappa\left(\rho^{\prime\prime}(r, t) + \frac{2}{r}\rho^{\prime}(r, t)\right) \\%
\end{equation}%
where a prime indicates the derivative with respect to $r$.

The non-equilibrium thermodynamic relations need to be supplied by the balance equations. For a one-component system there are four balance equations, for the density, momentum, energy and entropy. For a spherically symmetric fluid with all fluxes along the radial direction with the barycentric velocity having only the radial component $\velocity$. The balance equations are
\begin{equation}\label{eq/Non-equilibrium/05a}
\begin{array}{rl}
\displaystyle \frac{\partial\rho}{\partial t} &= \displaystyle - \frac{1}{r^2}\frac{\partial}{\partial r} %
\left( r^2 \rho\velocity \right) \\\\%
\displaystyle \frac{\partial\rho \velocity}{\partial t} &= \displaystyle - \frac{1}{r^2}\frac{\partial}{\partial r} %
\left( r^{2} (p_{n} + \rho\velocity^{2}) \right) - 2 \frac{p}{r} \\\\%
\displaystyle \frac{\partial\rho u}{\partial t} &= \displaystyle - \frac{1}{r^2}\frac{\partial}{\partial r} %
\left( r^{2} J_{e} \right) \\\\%
\displaystyle \frac{\partial\rho s}{\partial t} &= \displaystyle - \frac{1}{r^2}\frac{\partial}{\partial r} %
\left( r^{2} J_{s}\right) + \sigma_{s}\\%
\end{array}
\end{equation}%
where $\sigma_{s}$ is the entropy production, while $J_{e}$ and $J_{s}$ are the total energy and entropy flux respectively. It is convenient to introduce the mass flux $J_{m}$, the momentum flux $J_{p}$, the heat flux $J_{q}$ as 
\begin{equation}\label{eq/Non-equilibrium/05b}
\begin{array}{rl}
J_{m} &\equiv \rho\velocity \\\\%
J_{p} &\equiv p_{n} + \rho\velocity^{2}\\\\%
J_{q} &\equiv J_{e} - J_{m}( h + \velocity^{2}/2 ) \\%
\end{array}
\end{equation}%
where $h = u + p/\rho$ is the specific enthalpy and $u$ is the specific internal energy. In stationary states the left hand side of all the equations in \eqr{eq/Non-equilibrium/05a} is zero, so it takes the following form
\begin{equation}\label{eq/Non-equilibrium/05}
\begin{array}{rl}
\displaystyle \left( r^{2} J_{m}\right)^{\prime} &= 0 \\\\%
\displaystyle \left( r^{2} J_{e}\right)^{\prime} &= 0 \\\\%
\displaystyle \left( r^{2} J_{p}\right)^{\prime} &= 2 r p \\\\%
\displaystyle \left( r^{2} J_{s}\right)^{\prime} &= r^2 \sigma_{s} \\%
\end{array}
\end{equation}%
Note, that unlike in the case of planar interface, the mass flux and the energy flux in the direction across the interface are not constant. They decrease inversely proportionally to the radius squared. This leads to the fact that in stationary states both the mass flux and the energy flux  become infinite at the origin. In order to make this possible one has to introduce a source or sink for heat and mass at a spherical surface close to the center of the bubble. Using previously derived integral relations for the transfer coefficients for heat and mass transfer through a surface we only need equilibrium profiles. We therefore refrain from a further analysis of stationary states.

To obtain the expressions for the interfacial resistances, we need to consider non-equilibrium. For a proper description of a non-equilibrium process the Gibbs relation is required. Following \cite{Glavatskiy2009}, we write the Gibbs relations for a spherical system as
\begin{equation}\label{eq/Non-equilibrium/03}
T\,{\frac{ds}{dt}} = {\frac{du}{dt}} + p\,{\frac{dv}{dt}} -
v\,\velocity\,\frac{1}{r^2}\frac{\partial}{\partial r}\left(r^2\,\gamma_{rr}\right)
\end{equation}%
where $s$, $u$, $v \equiv 1/\rho$ are the specific entropy, internal energy and volume respectively, which are related to the other thermodynamic quantities in a manner, which is similar to \eqr{eq/Non-equilibrium/01} and \eqr{eq/Non-equilibrium/02}. Furthermore, $d/dt$ is the substantial (barycentric) time derivative: $d/dt = \partial/\partial t + \velocity\, \partial/\partial r$. Note, that \eqr{eq/Non-equilibrium/03} is not restricted to the stationary state condition. All the quantities depend in general both on position and time. However, the arguments $(r, t)$ were omitted to simplify the notation.

Combining the above equations we obtain the expression for the local entropy production in the interfacial region:
\begin{equation}\label{eq/Non-equilibrium/06}
\sigma_{s} = J_{q}\left(\frac{1}{T}\right)^{\prime}
\end{equation}%
The entropy production is always positive and therefore the heat flux is given by the linear constitutive relation
\begin{equation}\label{eq/Non-equilibrium/07}
\left(\frac{1}{T}\right)^{\prime} = r_{qq} J_{q}
\end{equation}%

In the context of the square gradient theory the local resistivity profile $r_{qq}$ is represented by the two terms\cite{bedeaux/vdW/I, glav/grad1}, a homogeneous term and a square gradient term:
\begin{equation}\label{eq/Non-equilibrium/08}
r_{qq}(r,t) = r_{qq,\,0}(T, \rho) + A(T, \rho) |\rho^{\prime}|^{2}
\end{equation}
where $\rho$ and $T$ also depend on position and time. In principle, the homogeneous term $r_{qq,\,0}$ is given by a kind of equation of state for the resistivity. Given the lack of knowledge about the temperature and the density dependence we model this term by a linear interpolation between two known values of the bulk resistivities:
\begin{equation}\label{eq/Non-equilibrium/09}
r_{qq,\,0}(\rho) = r_{qq}^{i} + (r_{qq}^{o}-r_{qq}^{i}){\frac{\rho-\rho^{i}}{\rho^{o}-\rho^{i}}}
\end{equation}
where $r_{qq}^{i}$ and $r_{qq}^{o}$ are the resistivities of the coexisting homogeneous inner and outer phase with a flat interface, taken for instance at the temperature of the outer boundary of the box. $\rho^i$ and $\rho^o$ are similarly densities of coexisting homogeneous inner and outer phases with a flat interface at this temperature. 

It was shown earlier \cite{Glavatskiy2010b} that the existence of the square gradient contribution is consistent with the second law of thermodynamics and gives a more accurate description of the interfacial resistances for the planar interface than the expressions of the kinetic theory. The coefficient $A$ in the square gradient contribution may depend on the local temperature and density. In the previous work for a planar interface \cite{glav/grad1} it was modeled as
\begin{equation}\label{eq/Non-equilibrium/10}
A = \alpha\,\frac{r_{qq}^{o}+r_{qq}^{i}}{\max[\nabla{\rho(\vR)}]^{2}}
\end{equation}
where $\alpha$ is a dimensionless coefficient of the order of unity and $\max[\nabla{\rho(\vR)}]$ is the maximum value of the density gradient for the planar interface. This maximum corresponds to the inflection point of the density profile at the temperature considered. 

We note that $r_{qq}^{i}$, $r_{qq}^{o}$, $\rho^{i}$, $\rho^{o}$, and $\max[\nabla{\rho(\vR)}]$ are parameters of the resistivity profile $r_{qq}(r,t)$. In the context of the square gradient model $r_{qq}$ depends only on $T(r,t)$, $\rho(r,t)$ and $\rho^{\prime}(r,t)$, the local values of the temperature, the density and the density gradient. Thus, the above parameters are just constants. In particular, these values do not depend on the curvature of the interface and should be calculated for a flat interface. A dependence of these parameters on the surface curvature would make the theory non-local. As this is inherently inconsistent with the square gradient description, we will not study this here. Note, that due to this, the values of $r_{qq,\,0}$ in the origin and at the outer boundary are not equal to $r_{qq}^{i}$ and $r_{qq}^{o}$ respectively. 

\section{Excess resistance of a spherical surface.}\label{sec/Excess}

On a macroscopic level the interfacial region is described by the so-called excess quantities. In equilibrium they allow one to consider the entire interface as a single entity. The use of excess quantities can be extended to non-equilibrium and this idea has been proven to be useful in many applications \cite{SKDB/surface}, showing that non-equilibrium interface can also be considered as a single entity.

Excess quantities are defined using local continuous profiles. A discussion of the technical details of this definition has been presented in \cite{Glavatskiy2009}. A general theory of the non-equilibrium interface in terms of the excess densities in curvilinear coordinates using the non-equilibrium local description was presented in \cite{Glavatskiy2010a}. Here we briefly summarize the main points for a spherical interface.

A key quantity in a macroscopic description of the interface is the excess of a thermodynamic density (mass density, energy density, entropy density, etc.). While the density is measured per unit of volume, the excess of a density is measured per unit of surface area. It depends on the position $R$ of so-called dividing surface, which may be chosen arbitrarily inside the interfacial region. It is one of the properties of the surface, that while this choice affects the values of different thermodynamic properties, it does not affect the thermodynamic relations. This is true in equilibrium and has been recently verified in non-equilibrium \cite{Savin2012, Glavatskiy2009, bedeaux/vdW/II}. For a thermodynamic density per unit of volume $\phi$ its excess in spherical coordinates $\widehat{\phi}$ is defined as
\begin{equation}\label{eq/Excess/01}
\widehat{\phi}(R) \equiv \frac{1}{R^2} \int_{0}^{L}{dr\,r^{2}\,\phi^{ex}(R,r)} 
\end{equation}
where $L$ is the radius of the spherical box, $R$ is the position of a dividing surface and 
\begin{equation}\label{eq/Excess/02}
\phi^{ex}(R,r) \equiv \phi(r,t) - \phi^{s,i}(r,t)\,\Theta(R-r) - \phi^{s,o}(r,t)\,\Theta(r-R)
\end{equation}
where $\Theta$ is the Heaviside function (1 for positive and 0 for negative values of the argument), while $\phi^{i}$ and $\phi^{o}$ are values of the homogeneous densities inside and outside the bubble extrapolated to the dividing surface $R$. For equilibrium profiles $\phi^{i}$ and $\phi^{o}$ are independent of $r$ and $t$ and we take $\phi^{i}$ to be equal to the value of $\phi$ in the center of the bubble and $\phi^{o}$ to be equal to the value of $\phi$ at the outer boundary.

There exist various choices of the dividing surface and a common one is the the equimolar dividing surface $R^{\rho}$. In case of a one-component fluid it is defined as $\widehat{\rho}(R^{\rho}) = 0$. The other choices of the dividing surface are the surface of tension and the inflection point. The further analysis does not depend on a particular choice of the dividing surface, and we will not specify it until the results. Each of the dividing surfaces can be used to define the size of the bubble. 

We will now consider stationary states. One of the relevant quantities for heat and mass transport across the interface is the excess entropy production $\widehat{\sigma}_{s}$. The local entropy production is a density per unit of volume, so the excess entropy production is defined using \eqr{eq/Excess/01}. It can be shown \cite{Glavatskiy2010a} that the excess of the local entropy production which is given by \eqr{eq/Non-equilibrium/06}, is
\begin{equation}\label{eq/Excess/03}
\begin{array}{rl}
\widehat{\sigma}_{s} %
= & \displaystyle 
  J_{q}^{i}\,\Delta\frac{1}{T}  %
- J_{m}\,\left(\Delta\frac{\widetilde{\mu}}{T} - \widetilde{h}^{i}\,\Delta\frac{1}{T}\right) \\\\
= & \displaystyle 
  J_{q}^{o}\,\Delta\frac{1}{T} %
- J_{m}\,\left(\Delta\frac{\widetilde{\mu}}{T} - \widetilde{h}^{o}\,\Delta\frac{1}{T}\right) \\
\end{array}
\end{equation}%
where $\Delta (1/T) \equiv 1/T^{o} - 1/T^{i} $ and $\Delta \widetilde{\mu}/T \equiv \widetilde{\mu}^{o}/T^{o} - \widetilde{\mu}^{i}/T^{i}$ are the jumps between the values of the corresponding functions extrapolated from the two bulk regions to the interfacial region and evaluated at the dividing surface. Furthermore, the superscripts $^{i}$ and $^{o}$ indicate the values of the corresponding homogeneous quantities inside and outside the bubble, which are extrapolated to the dividing surface $R$. All the quantities in \eqr{eq/Excess/03} depend on the choice of the dividing surface. Furthermore, $\widetilde{\mu} \equiv \mu + \velocity^2/2$ and $\widetilde{h} \equiv h + \velocity^2/2$.

Note, that in contrast to \eqr{eq/Non-equilibrium/06}, expression for the excess entropy production contains an additional term proportional to the total mass flux. This contribution to the excess entropy production of co-moving flux of matter is due to evaporation or condensation. It means that a difference in the temperature on two sides of the interface will cause not only the heat flux across the interface, but also the mass flux across the interface. Furthermore, \eqr{eq/Excess/03} for the excess entropy production contains two expressions, each one with a different heat flux at the dividing surface, $J_{q}^{i}$ and $J_{q}^{o}$. These fluxes are different, and their difference is determined by the enthalpy of the phase change: $J_{q}^{i} - J_{q}^{o} = J_{m}\left(\widetilde{h}^{o} - \widetilde{h}^{i}\right)$. 

The form of the excess entropy production \eqref{eq/Excess/03} suggests the force-flux relations
\begin{equation}\label{eq/Excess/04}
\begin{array}{rl}
\displaystyle \Delta\frac{1}{T} =& R_{qq} J_{q}^{\nu} - R^{\nu}_{qm} J_{m} \\\\
\displaystyle \Delta\frac{\widetilde{\mu}}{T} - \widetilde{h}^{\nu}\,\Delta\frac{1}{T}
= & R^{\nu}_{mq} J_{q}^{\nu} - R^{\nu}_{mm} J_{m} \\
\end{array}
\end{equation}%
where $\nu$ is either $i$ or $o$. Note, that $R_{qq}$ is independent of $\nu$. The coefficients $R_{qq}$, $R^{\nu}_{qm}$, $R^{\nu}_{mq}$ and $R^{\nu}_{mm}$ are the resistances of the interface to the heat and mass transfer. They determine the jumps of the temperature and the chemical potential across the bubble interface. Like all other interfacial quantities, these resistances depend on the size of the bubble. It is the aim of this paper to investigate this dependence.

In the context of linear irreversible thermodynamics these resistances are determined by equilibrium properties of the interface. Analysis in \cite{Glavatskiy2010a} and \cite{Glavatskiy2010b} applied to a one-component system gives the following expressions for the interfacial resistances:
\begin{equation}  \label{eq/Excess/05}
\begin{array}{rl}
R_{qq} & = \excess{r_{qq}}
\\\\
R^{\nu}_{qm} = R^{\nu}_{mq} & = \excess{r_{qm}^{\nu}}
\\\\
R^{\nu}_{mm} & =  \excess{r_{mm}^{\nu}} \\
\end{array}
\end{equation}
where $\excess{\phi}$ denotes the excess of a quantity which is not a density. For the resistances it is defined as
\begin{equation}\label{eq/Excess/06}
\excess{\phi}(R) \equiv R^2 \int_{0}^{L}{dr\,\frac{1}{r^2}\,\phi^{ex}(R,r)} 
\end{equation}
where $\phi^{ex}$ is still given by \eqr{eq/Excess/02} and $r_{qq}(r)$ is given by \eqr{eq/Non-equilibrium/08}. Furthermore, we have introduced as short hand notation:
\begin{equation}  \label{eq/Excess/07}
\begin{array}{rl}
r_{qm}^{\nu} & \equiv r_{qq}\,(h^{\nu}-h)
\\\\
r_{mm}^{\nu} & \equiv  r_{qq}\,(h^{\nu}-h)^2
\end{array}
\end{equation}
Note, that in contrast to the multicomponent system, these quantities are not independent: they are proportional to the heat resistivity $r_{qq}$. Note furthermore, that for equilibrium profiles $\widetilde{h} = h$. Furthermore, $h^{i}$ is equal to the actual value of $h$ in the center of the bubble and $h^{o}$ is equal to the actual value of $h$ at the outer boundary.

\section{Results and discussion.}\label{sec/Results}

We consider cyclohexane and use the van der Waals equation of state at the temperature $T = 330$ K. The van der Waals parameters $A = 2.195$ [J m$^3$/mol] and $B = 14.13 \times 10^{-5}$ [m$^3$/mol]. The parameter of the square gradient model $\kappa = 1.7282 \times 10^{-15}$ [J m$^5$/kg$^2$], which gives the value of the surface tension of the planar interface $\gamma  = 0.0370$ [N/m]. 

Fluid is placed in a spherical container with the radius $L = 80$ nm with a bubble being formed in the center. To avoid boundary effects, when the size of the bubble is close to the size of the container, the maximum bubble size considered is equal to 65 nm. As it was mentioned above, there exist a minimum size of the bubble, due to the finite compressibility of the liquid. For cyclohexane this size is equal to approximately 18 nm. To avoid effects of instability, the minimum bubble size considered here is equal to 25 nm. This range of bubble sizes is on the one hand good to consider large curvatures, and on the other hand it gives sufficient data to extrapolate them to planar interfaces. 

\begin{figure}[hbt!]
\centering
\includegraphics[scale=\scaleprofile]{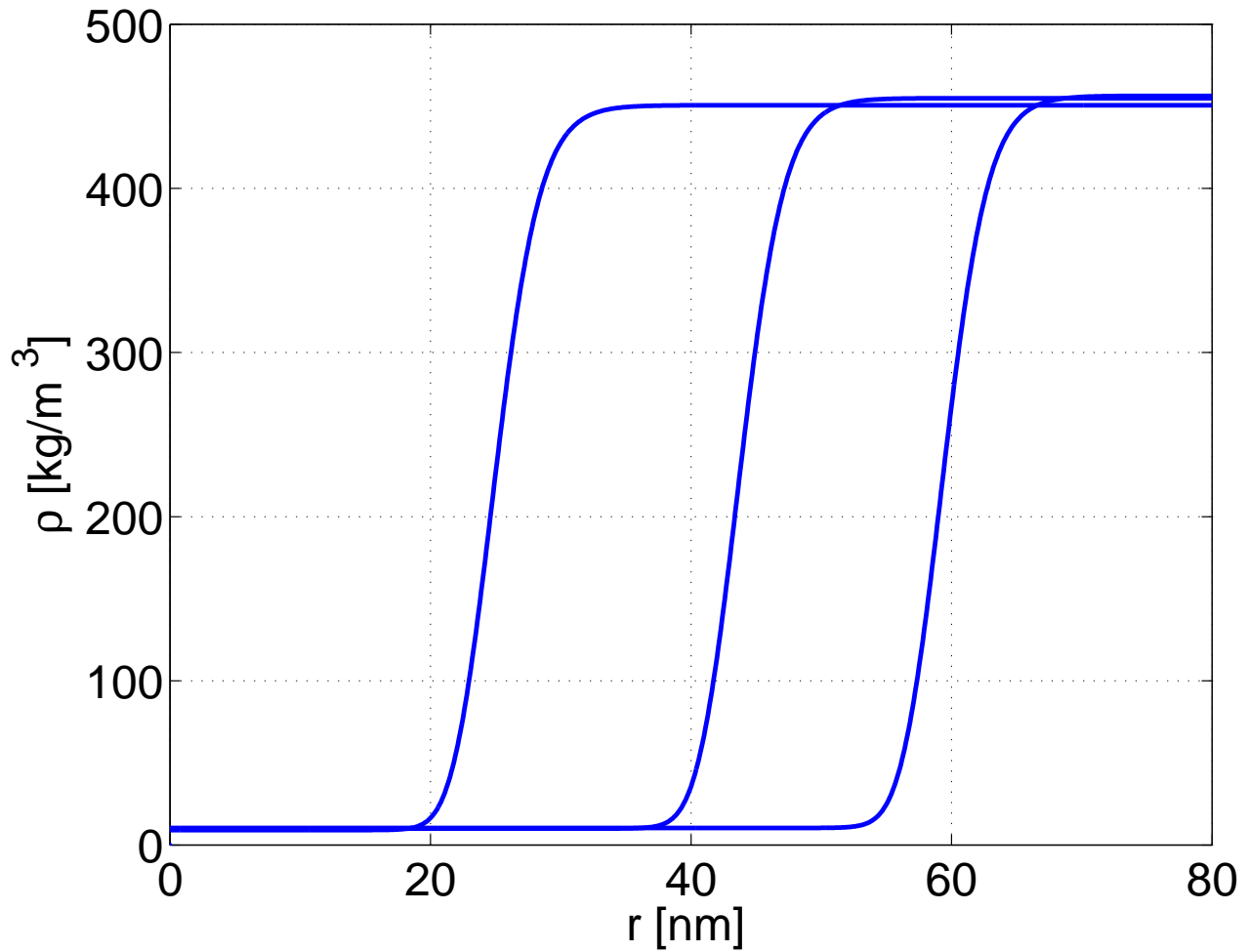}
\caption{Density profiles for various bubble sizes.}\label{fig/rho(r)}
\end{figure}

The typical profiles of the density are given in \figr{fig/rho(r)}. The curves in \figr{fig/rho(r)} represent the bubbles of different size. Gradual increase of the total mass of the fluid leads to a gradual decrease of the bubble radius. 

\begin{figure}[hbt!]
\centering
\includegraphics[scale=\scaleprofile]{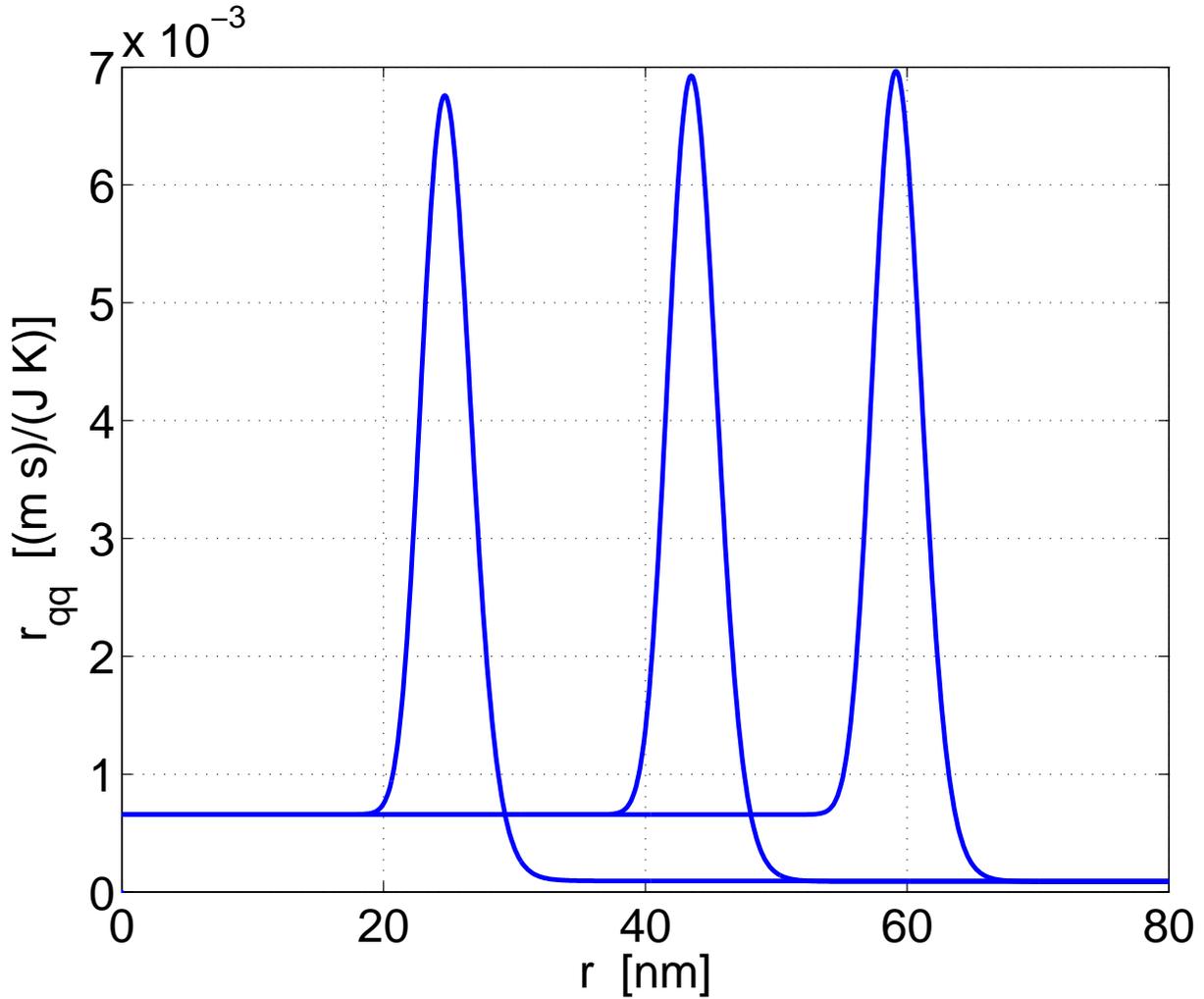}
\caption{Resistivity profiles for various bubble sizes.}\label{fig/r_qq(r)}
\end{figure}

The local resistivity profiles which are modeled by \eqr{eq/Non-equilibrium/08} from these density profiles are presented in \figr{fig/r_qq(r)}. \eqr{eq/Non-equilibrium/08} contains one parameter $\alpha_{qq}$, which determines the significance of the square gradient contribution to the local resistivity and therefore the magnitude of the peak in the interfacial region. It was shown in \cite{Glavatskiy2010b} that in order to satisfy the second law of thermodynamics this parameter should differ from zero. In this calculations we used the value $\alpha_{qq} = 9$. The thermal conductivity of the gas and the liquid phase $\ell^{i}_{qq} = 0.0140$ [W/(m K)] and $\ell^{o}_{qq} = 0.1130$ [W/(m K)]. The resistivities of these phases are calculated as $r_{qq}^{\nu} = (\ell^{\nu}_{qq}T^{2})^{-1}$.


\begin{figure}[hbt!]
\centering
\includegraphics[scale=\scaleradius]{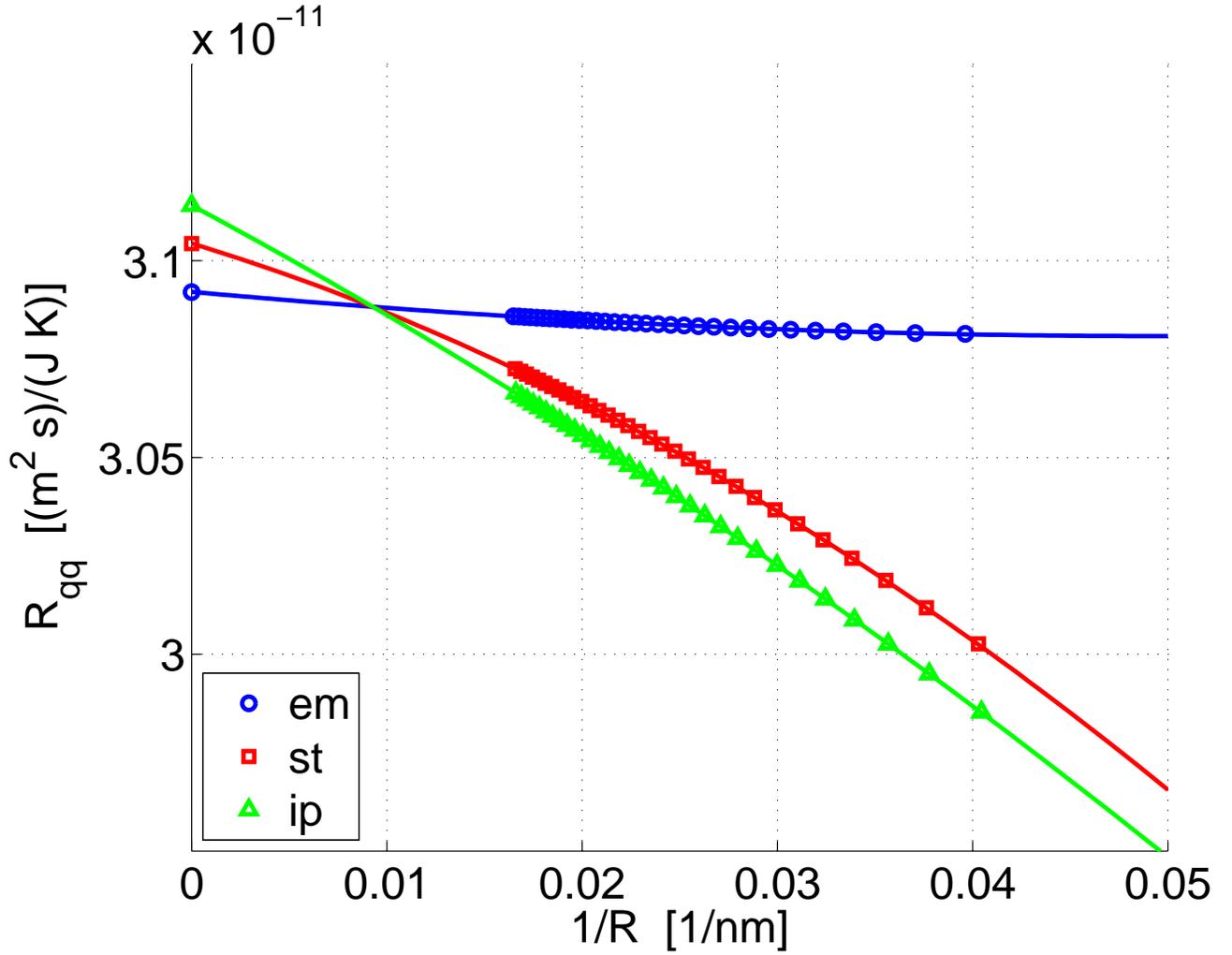}
\caption{Excess resistance $R_{qq}$ as a function of the bubble curvature for different dividing surfaces: equimolar surface (em), surface of tension (st), inflection point (ip). Symbols represent the data from \eqr{eq/Excess/05}, lines represent the quadratic fit \eqr{eq/Results/01}.}\label{fig/Rqq_m(R)}
\end{figure}

\begin{figure}[hbt!]
\centering
\includegraphics[scale=\scaleradius]{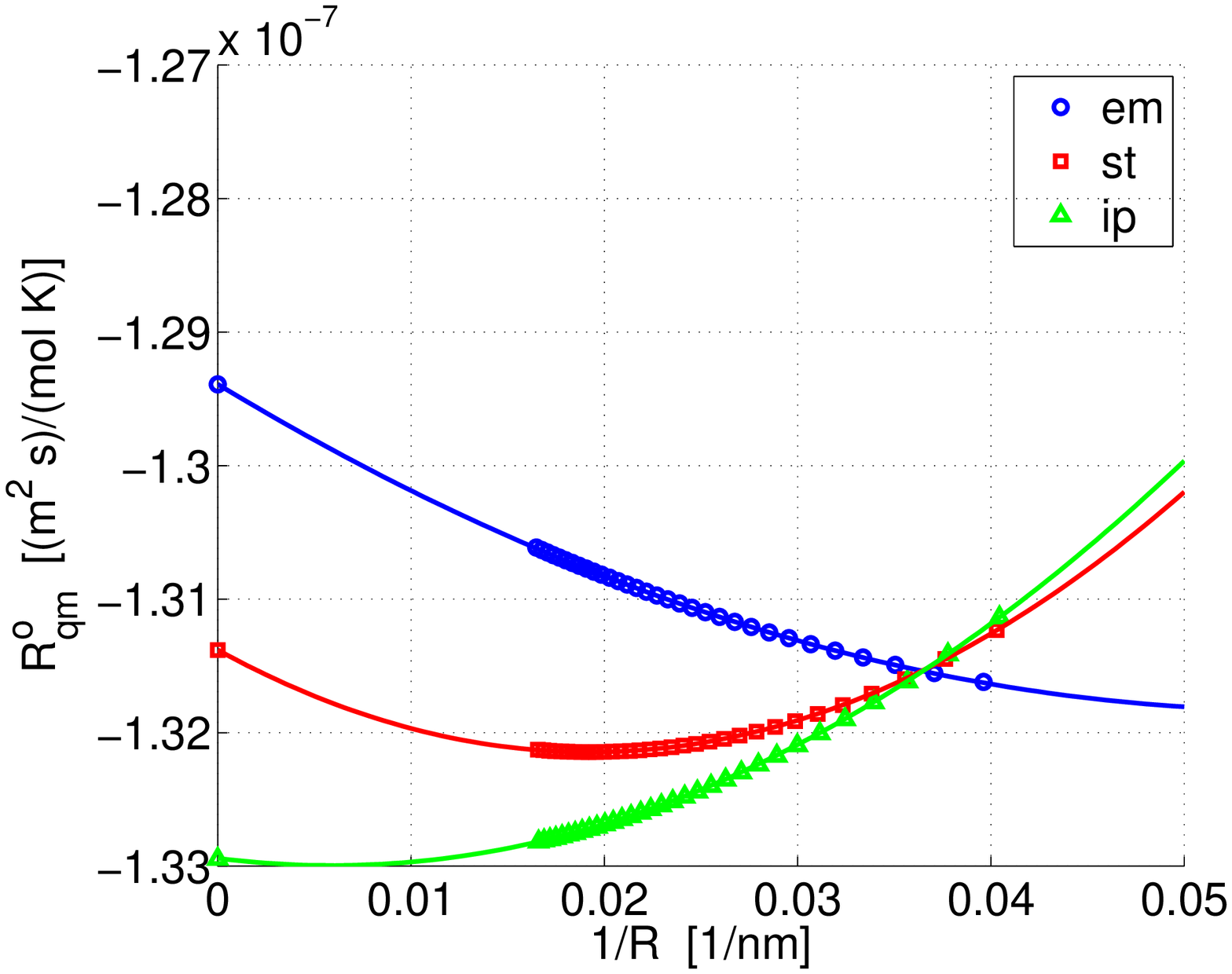}
\caption{Excess resistance $R_{qm}^{o}$ as a function of the bubble curvature for different dividing surfaces: equimolar surface (em), surface of tension (st), inflection point (ip). Symbols represent the data from \eqr{eq/Excess/05}, lines represent the quadratic fit \eqr{eq/Results/01}.}\label{fig/Rqm_m(R)}
\end{figure}

\begin{figure}[hbt!]
\centering
\includegraphics[scale=\scaleradius]{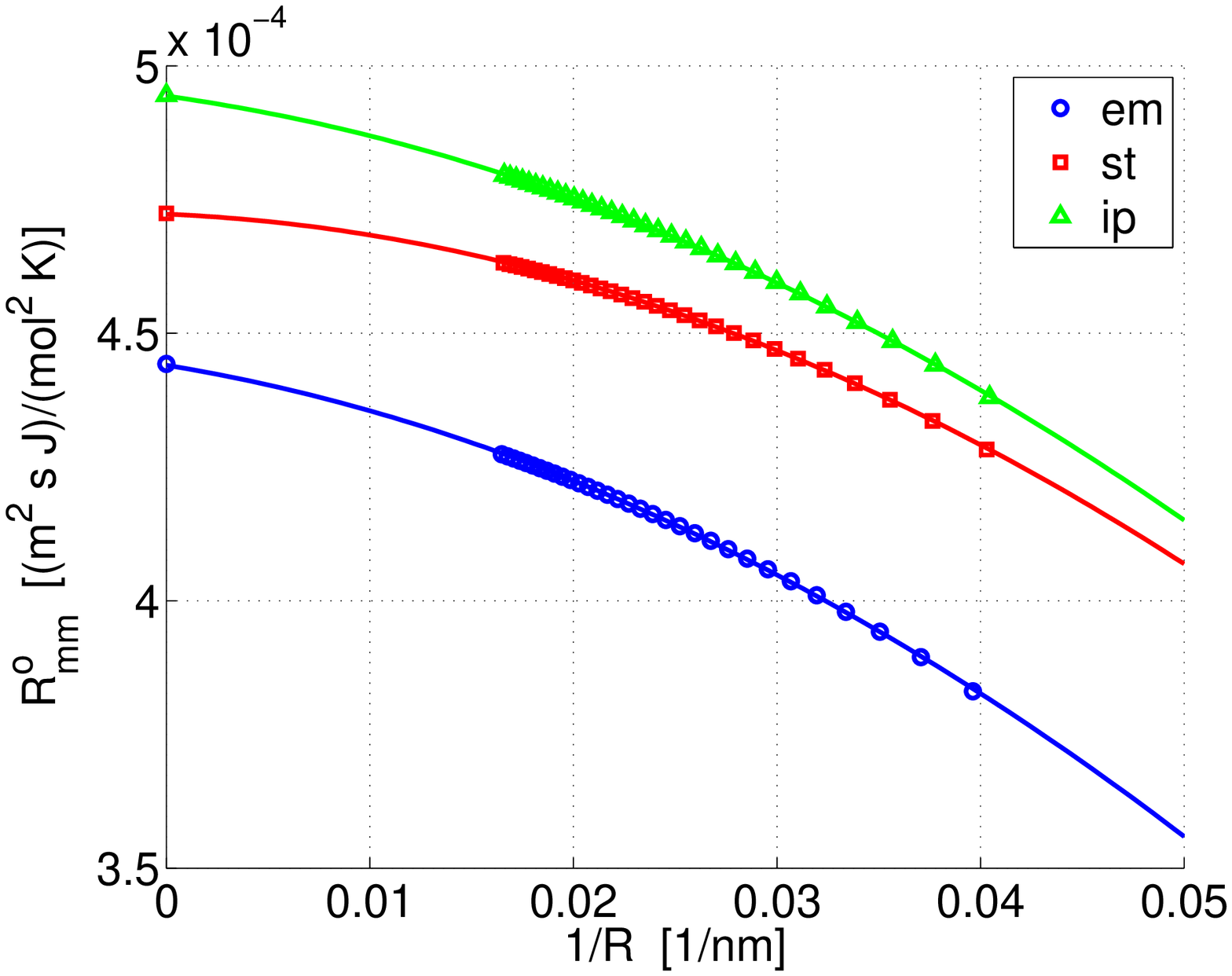}
\caption{Excess resistance $R_{mm}^{o}$ as a function of the bubble curvature for different dividing surfaces: equimolar surface (em), surface of tension (st), inflection point (ip). Symbols represent the data from \eqr{eq/Excess/05}, lines represent the quadratic fit \eqr{eq/Results/01}.}\label{fig/Rmm_m(R)}
\end{figure}

Next we consider the dependence of the interfacial resistances $R_{qq}$, $R^{o}_{qm}$ and $R^{o}_{mm}$ on the curvature. We do this for three choices of the dividing surface: equimolar surface (em), surface of tension (st) and the inflection point (ip). The dependencies are presented in \figr{fig/Rqq_m(R)}, \figr{fig/Rqm_m(R)} and \figr{fig/Rmm_m(R)} respectively. In addition, the value of the corresponding resistances for the planar interface is indicated (the point of zero curvature). 


Furthermore, a quadratic fit of the form
\begin{equation}  \label{eq/Results/01}
R_{ab} = \displaystyle R_{ab,\,0}\,\left(1 + R_{ab,\,1}\,\frac{1}{R} + R_{ab,\,2}\,\frac{1}{R^2}\right)\\\\
\end{equation}
where $ab$ stands for either $qq$, or $qm$, or $mm$, is applied to the data and plotted by a solid line. The fit includes the zero-curvature value of the resistances. The values of the coefficients are given in \tblr{tbl/Rqq}, \tblr{tbl/Rqm} and \tblr{tbl/Rmm}. In addition, the actual value for the planar resistance is given.

\begin{center}
\begin{longtable}{c @{\;\qquad} r @{\;\qquad} r  @{\;\qquad} r @{\;\qquad}r  @{\;\qquad}}
\caption{The values of the $R_{qq}$ resistance for the planar interface and the coefficients of the quadratic fit \eqref{eq/Results/01} for different dividing surfaces} \label{tbl/Rqq} \\
\hline
dividing & $R_{qq,\,\infty}$,   & $R_{qq,\,0}$, & $R_{qq,\,1}$, & $R_{qq,\,2}$, \\
surface & (m$^2$ s)/(J K)   & (m$^2$ s)/(J K) & nm & nm$^2$ \\
\hline
em & 3.0920 $\times 10^{-11}$ &  3.0921 $\times 10^{-11}$ 	& - 0.1469  	& 1.4838  \\
st & 3.1044 $\times 10^{-11}$ &  3.1044 $\times 10^{-11}$  	& - 0.4852 		& - 8.1855 \\
ip & 3.1146 $\times 10^{-11}$ &  3.1141 $\times 10^{-11}$  	& - 0.8534    	& - 4.2083  \\
\hline
\end{longtable}%
\end{center}
\begin{center}
\begin{longtable}{c @{\;\qquad} r @{\;\qquad} r  @{\;\qquad} r @{\;\qquad}r  @{\;\qquad}}
\caption{The values of the $R_{qm}$ resistance for the planar interface and the coefficients of the quadratic fit \eqref{eq/Results/01} for different dividing surfaces} \label{tbl/Rqm} \\
\hline
dividing & $R_{qm,\,\infty}$,   & $R_{qm,\,0}$,  & $R_{qm,\,1}$, & $R_{qm,\,2}$, \\
surface & (m$^2$ s)/(mol K)   & (m$^2$ s)/(mol K) & nm & nm$^2$ \\
\hline
em & -1.2939 $\times 10^{-7}$ &  -1.2939 $\times 10^{-7}$ 	& 0.6758  	& -  6.0481  \\
st & -1.3138 $\times 10^{-7}$ &  -1.3138 $\times 10^{-7}$  	& 0.6040 	& - 15.6789 \\
ip & -1.3294 $\times 10^{-7}$ &  -1.3294 $\times 10^{-7}$  	& 0.1385    & - 11.7218  \\
\hline
\end{longtable}%
\end{center}
\begin{center}
\begin{longtable}{c @{\;\qquad} r @{\;\qquad} r  @{\;\qquad} r @{\;\qquad}r  @{\;\qquad}}
\caption{The values of the $R_{mm}$ resistance for the planar interface and the coefficients of the quadratic fit \eqref{eq/Results/01} for different dividing surfaces} \label{tbl/Rmm} \\
\hline
dividing & $R_{mm,\,\infty}$,  & $R_{mm,\,0}$, & $R_{mm,\,1}$, & $R_{mm,\,2}$\\
surface & (m$^2$ s J)/(mol$^2$ K)   & (m$^2$ s J)/(mol$^2$ K) &  nm & nm$^2$\\
\hline
em & 4.4428 $\times 10^{-4}$ &  4.4402 $\times 10^{-4}$ 	& - 1.4042  	& - 51.2815  \\
st & 4.7242 $\times 10^{-4}$ &  4.7227 $\times 10^{-4}$  	& - 0.3404 		& - 48.5718  \\
ip & 4.9452 $\times 10^{-4}$ &  4.9435 $\times 10^{-4}$  	& - 1.0753   	& - 42.6580  \\
\hline
\end{longtable}%
\end{center}
%


It is interesting to observe, that the resistance curves for the different dividing surfaces have a common point of intersect. It is easy to understand that there could exist such point $R^{*}$, which we will call the \textit{static point}. This is the point where the resistance is the same for different dividing surfaces. In other words, at this point the excess resistance does not change when we change the dividing surface. For a resistance $R_{ab}$ which depends on the position $R$ of the dividing surface this condition is expressed as $dR_{ab}/dR = 0$. Using \eqr{eq/Excess/06} this gives the condition for the static point
\begin{equation}  \label{eq/Results/02}
\frac{dR_{ab}}{dR} = r^{o} - r^{i} + \frac{2}{R^{*}}\,R_{ab}(R^{*}) = 0
\end{equation}
The position of the static point is determined by the value of the difference between the bulk resistivities $r^{o} - r^{i}$ and the value of the excess resistance. The heat resistivity $r_{qq}$ of the gas phase is higher than the one of the liquid phase, so that $r_{qq}^{o} < r_{qq}^{i}$. Furthermore, $R_{qq}$ resistance is always positive. This makes the static point for $R_{qq}$ resistance to be positive. For the system studied, the static point is situated at approximately 107.7 nm, giving the value of $R_{qq}$ excess resistance approximately 3.0885$\times 10^{-11}$ (m$^3$ s)/(J K). $r_{qm}^{o}$ resistivity has a higher value for the liquid than for the gas, so that $r_{qm}^{o} > r_{qm}^{i}$. Furthermore, $R_{qm}^{o}$ resistance is always negative. This makes the static point for $R_{qm}^{o}$ resistance to be positive as well. For the system studied, the static point is situated at approximately  27.4 nm, giving the value of $R_{qm}^{o}$ excess resistance approximately -1.3155$\times 10^{-7}$ (m$^2$ s)/(mol K). $r_{mm}^{o}$ resistivity has a lower value for the outer phase than for the inner phase, so that $r_{mm}^{o} < r_{mm}^{i}$. $R_{mm}^{o}$ is also positive. Extrapolation of the data indicates that $R_{mm}^{o}$ changes the sign at approximately 9 nm, where a stable bubble does not exist. Within the domain of these curvatures $r^{i} - r^{o} \approx 1.4 \times 10^{5}$ (m s J)/(mol$^2$ K) is always larger than $2R_{mm}^{o}/R$, which makes \eqr{eq/Results/02} to have no solution for $R_{mm}^{o}$. This means that the static point for $R_{mm}$ resistance does not exist.


We also note, that the resistances do not necessarily depend monotonously on the curvature. While the heat resistance for the surface of tension and the inflection point decrease monotonously with increasing curvature, the heat resistance for the equimolar surface has a minimum value of approximately 3.0812$\times 10^{-11}$ (m$^2$ s)/(J K) when the size of the bubble is approximately 21.7 nm, which corresponds to the curvature 0.046 nm$^{-1}$. In order to understand this behavior, it is useful to consider the expression \eqref{eq/Non-equilibrium/08} for the local heat resistivity. For the equimolar surface the excess of the first term is equal to zero, $\excess{r_{qq,\,0}}(R^{em}) = 0$. Thus, excess of the heat resistance is entirely due to the square gradient contribution. It is the combination of contribution from the $A|\rho^{\prime}|^{2}$ factor and the $(R/r)^{2}$ factor to the excess, which makes the curvature dependence of the heat resistance for the equimolar surface to have a convex shape. The heat resistance for the other dividing surfaces, surface of tension and inflection point, has additional terms. Indeed, the change $\Delta R_{ab}$ of the resistance due to the change $\delta$ of the dividing surface is, according to \eqr{eq/Excess/06}
\begin{equation}  \label{eq/Results/03}
\Delta R_{ab} \approx \frac{dR_{ab}}{dR}\,\delta = \left(r^{o} - r^{i} + \frac{2}{R}\,R_{ab}(R)\right)\,\delta
\end{equation}
which increases with the curvature. Thus, the resistance for the surface of tension or the inflection point will diverge from the resistance for the equimolar surface when the curvature is increasing. We observe exactly this behavior in \figr{fig/Rqq_m(R)}, \figr{fig/Rqm_m(R)} and \figr{fig/Rmm_m(R)}.

\section{Conclusions.}\label{sec/Conclusion}

We have presented a framework to calculate the interfacial heat and mass resistances of curved surfaces. The method of determining interfacial resistances is in the context of the Gibbs excess quantities. In particular, the resistances are represented as the excesses of local resistivity profiles. Local resistivities are calculated with the help of the square gradient model, an approach which has shown to be useful for the description of the interfaces. 

Calculation of the interfacial resistances requires only equilibrium information about the system. In particular, the local resistivity profiles, which are the input quantities for the calculation of the excesses, are calculated with the help of the equilibrium density profiles. 

We have investigated how the interfacial resistances depend on the interface curvature. It was shown that they change with the curvature at least quadratically. In a closed system there exist restrictions on the minimum size of a stable bubble \cite{GlavatskiyRegueraBedeaux} because of the non-zero compressibility of the liquid. Thus, the curvature of a stable bubble in a closed system has an upper bound, which limits the magnitude of the resistance. In open systems, even though all bubbles and droplets are unstable, there is no restriction on the nucleus size \cite{Blockhuis1995}, so the excess resistance is not limited. However, when the curvature of the system becomes extremely high, the interfacial region fuses with the inner phase and the notion of the excess resistance is undefined. Further research is needed to address such high curvatures.

We have found, that the resistances for different dividing surfaces are different. The interfacial resistance cannot be measured on their own without specifying the dividing surface position. This behavior of the interfacial resistances is analogous to the fact that most of the Gibbs excess densities depend on the choice of the dividing surface\cite{Glavatskiy2009}. However, the form of the force-flux relations \eqref{eq/Excess/04} is the same for all choices of the dividing surface, just like the Euler relation between the Gibbs excess densities is the same for all choices of the dividing surface.

\acknowledgements

Dick Bedeaux wants to thank {\O}ivind Wilhelmsen for extensive discussions.

%

\bibliographystyle{unsrt}

\end{document}